\documentclass[aps,pra,twocolumn,english,10pt]{revtex4-1}
\usepackage{amsmath}
\usepackage{amssymb}
\usepackage{multirow}
\usepackage{graphicx}
\usepackage{babel}
\usepackage{natbib}
\usepackage{siunitx}
\usepackage{color}

\begin{document}

\title{Effects of intrinsic defects and alloying with Fe on the half-metallicity of Co$_2$MnSi}

\author{G. G. Baez Flores, I. A. Zhuravlev, and K. D. Belashchenko}

\affiliation{Department of Physics and Astronomy and Nebraska Center for Materials and Nanoscience, University of Nebraska-Lincoln, Lincoln, Nebraska 68588, USA}

\date{\today}

\begin{abstract}
The electronic structure and half-metallic gap of Co$_{2}$MnSi in the presence of crystallographic defects, partial Fe substitution for Mn, and thermal spin fluctuations are studied using the coherent potential approximation and the disordered local moment method. In the presence of 5\% Co or Mn vacancies the Fermi level shifts down to the minority-spin valence-band maximum. In contrast to NiMnSb, both types of Mn antisite defects in Co$_{2}$MnSi are strongly exchange-coupled to the host magnetization, and thermal spin fluctuations do not strongly affect the half-metallic gap. Partial substitution of Mn by Fe results in considerable changes in the Bloch spectral function near the Fermi level, which strongly deviate from the rigid-band picture. In particular, a light band with the Fe character crosses the Fermi level at about 50\% concentration. At room temperature, Fe substitution of up to 30\% slightly increases the spin polarization at the Fermi level.
\end{abstract}

\maketitle

\section{Introduction}

Half-metallic ferromagnets have electronic bands in only one spin channel at the Fermi level \cite{hm_groot,spinpol_fang}, which makes them attractive for applications in spintronic devices, such as magnetic tunnel junctions and spin valves \cite{Sakuraba2006,PALMSTROM1,GRAF20111}.
Co$_{2}$MnSi is a full-Heusler alloy with a high Curie temperature $T_C=\SI{985}{\kelvin}$ \cite{TC_CMS2,TC_CMS}, and both density-functional and GW calculations predict it to be a half-metal with a minority-spin gap of $\SI{0.82}{\eV}$ and $\SI{0.95}{\eV}$, respectively \cite{GWBANDS}. Spin polarization of 93\% at room temperature was observed in Co$_{2}$MnSi thin films in ultraviolet photoemission measurements \cite{Jourdan2014}. The half-metallic gap and the spin polarization can be sensitive to the presence of crystallographic defects. The effects of point defects on the half-metallic gap in full and half-Heusler alloys were studied using \textit{ab initio} calculations \cite{HalfHeuslerdefects,structuraldefects}, which showed that Co$_2$MnSi remains half-metallic at zero temperature in the presence of four types of defects: Mn$_{\textrm{Co}}$ and Mn$_{\textrm{Si}}$, Co vacancies, and Mn vacancies. Alloying with Fe on the Co sublattice also maintains the half-metallic gap at zero temperature \cite{CMFS_defects}.

For practical applications, it is important to consider the thermal stability of the half-metallic gap at room temperature \cite{Lezaic,NiMnSbpaper}. For example,
for the half-Heusler NiMnSb alloy it was found \cite{NiMnSbpaper} that the spins of Mn$_{\textrm{Sb}}$ antisites are weakly coupled to the host magnetization and are, therefore, easily disordered at low temperatures. As a result, the electronic states near the valence band maximum (VBM) are strongly broadened and shifted up toward the Fermi level, which leads to a strong depolarization of NiMnSb far below room temperature. One goal of the present paper is to examine the role of thermal spin fluctuations in Co$_2$MnSi with intrinsic point defects.

The spin polarization of a half-metal is expected to be more robust if the Fermi level lies close to the middle of the half-metallic gap. Because the Fermi level in Co$_2$MnSi lies closer to the VBM \cite{GWBANDS,Chadov_2009,EXP_DOS_CMFS}, partial substitution of Fe for Mn was explored as a way to shift the Fermi level upward and enhance the magnetoresistive effects \cite{Fermitailoring_Fe,CFS_MTJ_TMR,CMS_E_struc,GMR_CMFS_MTJ}. However, first-principles calculations, including those in the GW approximation \cite{GWBANDS}, show that the band structures of Co$_2$MnSi and Co$_2$FeSi differ by more than a rigid band shift. The second goal of this paper is to study the band structure of Co$_2$Mn$_{1-x}$Fe$_x$Si (CMFS) and (Co$_{1-x}$Fe$_x$)$_2$MnSi alloys as a function of Fe concentration and temperature.

\section{Computational methods}

First-principles calculations were performed within the atomic sphere approximation (ASA) in the Green's function-based tight-binding linear muffin-tin orbital (GF-LMTO) method \cite{ASA,Turek}, and substitutional alloying was described within the coherent potential approximation (CPA) \cite{GFLMTO}. The exchange and correlation were described within the generalized gradient approximation (GGA) \cite{GGA}. We used the experimental lattice parameter of Co$_2$MnSi (\SI{5.645}{\angstrom} \cite{CMSmagcurve}) in all calculations. Note that the lattice parameter of Co$_{2}$FeSi (\SI{5.64}{\angstrom} \cite{CFS_apar}) is very similar.

To minimize the discrepancies of the ASA calculations with the reference band structure of Co$_{2}$MnSi obtained using a full-potential linear augmented plane wave calculation \cite{FLEUR}, the atomic sphere radii were optimized \cite{note-radii} and the magnetic part of the GGA exchange-correlation field was scaled by a factor 1.15.

The effects of thermal spin disorder are introduced using the vector disordered local moments (DLM) model \cite{Gyorffy_1985,DLM_implementation,DLM_implementation2,DLM_CPA}, which assumes that the orientation of each local magnetic moment $i$ has an independent distribution function, $p_i(\theta_i)\propto \exp(\alpha_{i}\cos\theta_i)$, where $\theta_i$ is the polar angle made by that local moment with the global magnetization. The parameters $\alpha_i$ are different for inequivalent lattice sites and for different components on the same site; they are obtained from the solution of the coupled equations of the mean-field approximation (MFA) at the given temperature with the magnetic exchange parameters calculated using the linear response technique \cite{LIECHTENSTEIN}.

\begin{figure}[h]
    \centering
    \includegraphics[width=0.9\columnwidth]{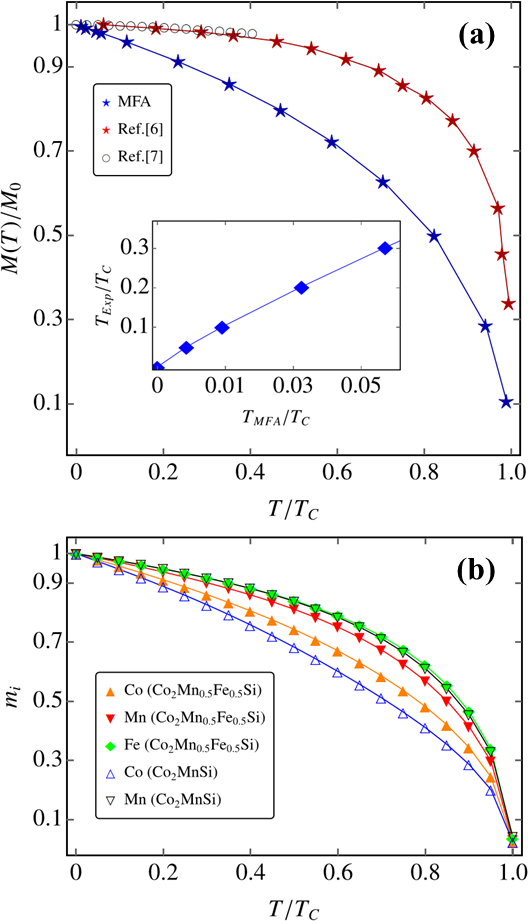}
    \caption{(a) Magnetization curve for Co$_{2}$MnSi compared with experimental data \cite{TC_CMS,TC_CMS2}. Inset: conversion from $T^\prime$ to $T$ using $M_\mathrm{MFA}(T')=M_{exp}(T)$. (b) Reduced magnetization for each magnetic sublattice in Co$_{2}$MnSi (empty symbols) and Co$_{2}$Mn$_{0.5}$Fe$_{0.5}$Si (filled symbols).}
    \label{fig:MT}
\end{figure}

Because the decline of the magnetization at low temperatures is too fast in MFA, we use the experimental $M_{exp}(T)$ curve for Co$_{2}$MnSi \cite{CMSmagcurve} to improve the predictions for the electronic structure. Specifically, to predict the electronic properties at temperature $T$, we determine the temperature $T'$ such that MFA with the calculated exchange parameters gives the magnetization $M_\mathrm{MFA}(T')=M_{exp}(T)$; this conversion is illustrated in Fig. \ref{fig:MT}(a). The electronic structure obtained using the DLM method at temperature $T'$ is then reported as corresponding to temperature $T$.
\section{Pure C\lowercase{o}$_{2}$M\lowercase{n}S\lowercase{i} and C\lowercase{o}$_{2}$F\lowercase{e}S\lowercase{i}}

\begin{figure}[h]
    \centering
    \includegraphics[width=0.9\columnwidth]{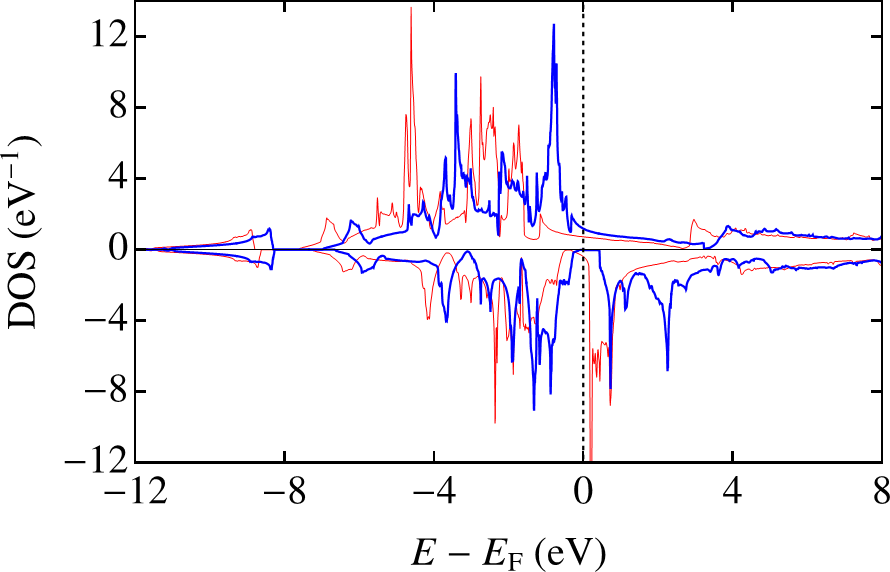}
    \caption{Density of states for majority (positive scale) and minority-spin (negative scale) electrons in Co$_2$MnSi (blue) and Co$_2$FeSi (red).}
    \label{fig:CMS_DOS}
\end{figure}

Spin-resolved densities of states (DOS) in Co$_2$MnSi and Co$_2$FeSi are shown in Fig.\ \ref{fig:CMS_DOS}, and the minority-spin band structures in Fig. \ref{fig.defects}(a-b). Co$_2$MnSi is seen to be a half-metal with a saturated magnetic moment of 5 $\mu_B$ and a half-metallic gap of $\SI{0.53}{\eV}$ in GF-LMTO, which is underestimated compared to the full-potential value of $\SI{0.82}{\eV}$ \cite{GWBANDS}. The half-metallic band gap is indirect and corresponds to the $\Gamma\rightarrow{\textrm{X}}$ transition energy. The direct gap at the $\Gamma$ point is wider by $\SI{6}{\milli\eV}$. On the other hand, Co$_2$FeSi is a ferromagnetic metal with a magnetic moment of 5.81 $\mu_B$ in agreement with full-potential calculations \cite{GWBANDS}. The dispersive band with the Fe character, which was noted in Ref.\ \onlinecite{GWBANDS}, forms an electron pocket around the X point and is clearly seen in Fig.\ \ref{fig.defects}b.
\section{Vacancies and antisites in C\lowercase{o}$_{2}$M\lowercase{n}S\lowercase{i}}

\begin{figure*}[htb]
	\includegraphics[width=0.95\textwidth]{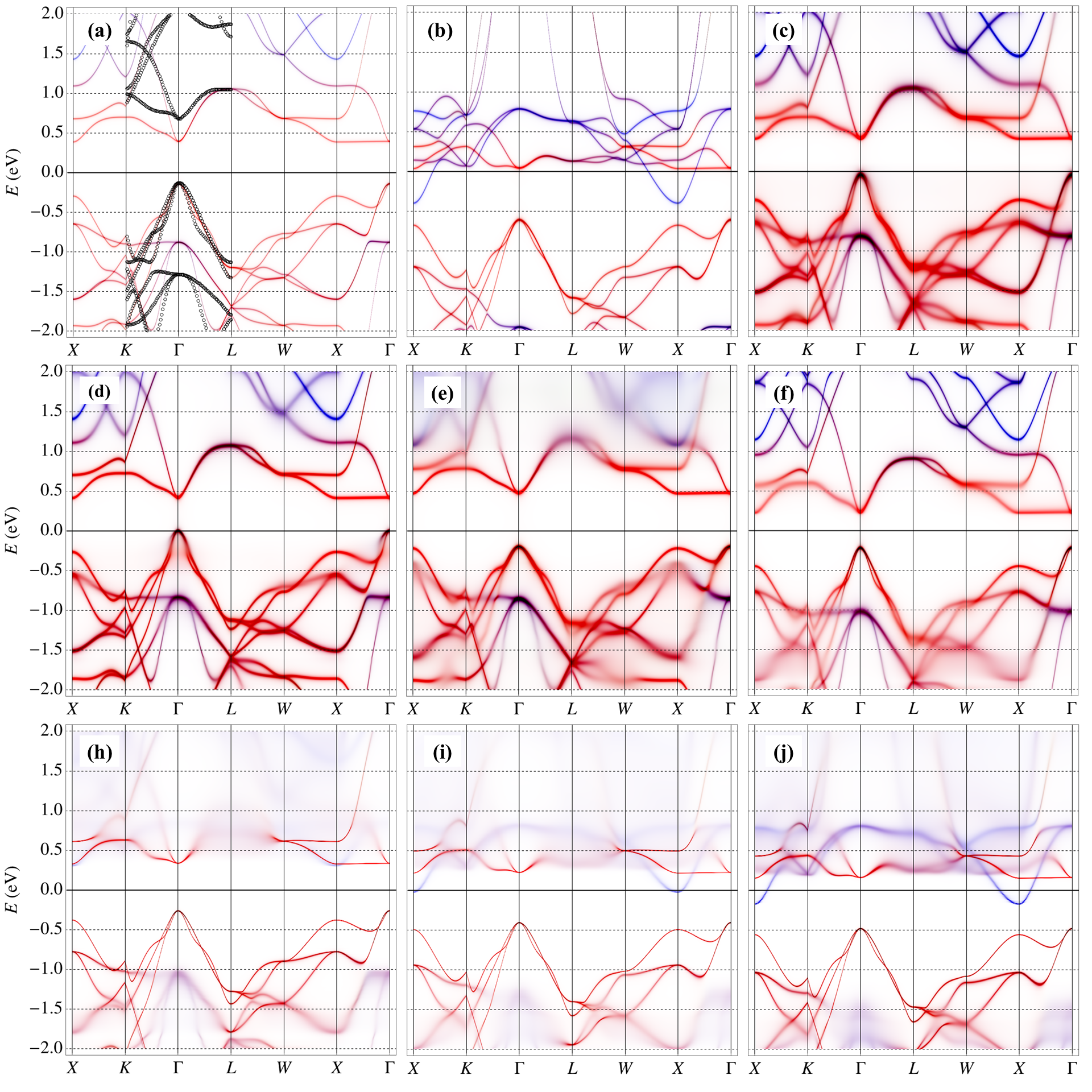}
	\caption{Minority-spin Bloch spectral functions at $T=0$ for (a) Co$_2$MnSi, (b) Co$_2$FeSi, (c-f) Co$_2$MnSi with defects, including (c) 5\% Co vacancies, (d) 5\% Mn vacancies, (e) 15\% Mn$_{\textrm{Si}}$ antisites, (f) 15\% Mn$_{\textrm{Co}}$ antisites, (h-j) Co$_{2}$Mn$_{1-x}$Fe$_{x}$Si at (h) $x=0.28$, (i) $x=0.55$, and (j) $x=0.72$. Color reflects the spectral character Co (red), Mn and Fe (blue), and Si (green). Black circles in panel (a) show the full-potential band structure. The Fermi level is at $E=0$.}
	\label{fig.defects}
\end{figure*}

In this section we consider the spectroscopic signatures of Co and Mn vacancies (V$_\mathrm{Co}$ and V$_\mathrm{Mn}$), Mn$_{\textrm{Si}}$ and Mn$_{\textrm{Co}}$ antisites, all of which preserve the half-metallic gap at zero temperature \cite{structuraldefects}.
The concentrations were set to 5\% and 15\% for vacancies and antisites, respectively, and the resulting minority-spin Bloch spectral functions for Co$_2$MnSi with each type of  defect are shown in Fig.\ \ref{fig.defects}(c-d).

The broadening of the electronic bands induced by the considered defects is fairly small even at the chosen large concentrations, and most of the bands close to the Fermi level remain well-defined. The spectral functions of Co$_{2}$MnSi with 1\% Co or Mn vacancies (not shown) are virtually indistinguishable from those of pure Co$_{2}$MnSi. Note that a similar conclusion was reached for Ni and Mn vacancies in half-Heusler NiMnSb \cite{HalfHeuslerdefects}. The conduction bands are almost unaffected by vacancies even at the 5\% concentration. However, vacancies decrease the minority spin-flip gap, i.e., the distance between the minority-spin VBM and the Fermi level, which in our calculation goes to zero at the vacancy concentration of about 5\%, as seen in Fig.\ \ref{fig.defects}(c-d). Thermal spin fluctuations at room temperature are expected to amplify the effect of this band shift on the spin polarization. Thus, large concentrations of Co or Mn vacancies can strongly reduce the spin polarization of Co$_2$MnSi.

As seen in Fig.\ \ref{fig.defects}(e-f), Mn$_{\textrm{Si}}$ and Mn$_{\textrm{Co}}$ antisites preserve the half-metallic gap even at a large concentration of 15\%, in agreement with Ref.\ \onlinecite{structuraldefects}. Mn$_{\textrm{Si}}$ antisites increase the half-metallic gap to \SI{0.67}{\eV} but have almost no effect on the spin-flip gap. In contrast, Mn$_{\textrm{Co}}$ antisites reduce the half-metallic gap while bringing the Fermi level closer to its center. Some of the occupied bands are considerably broadened in the presence of Mn$_{\textrm{Si}}$ or Mn$_{\textrm{Co}}$ antisites.

\begin{figure*}[t]
	\centering
	\includegraphics[width=0.95\textwidth]{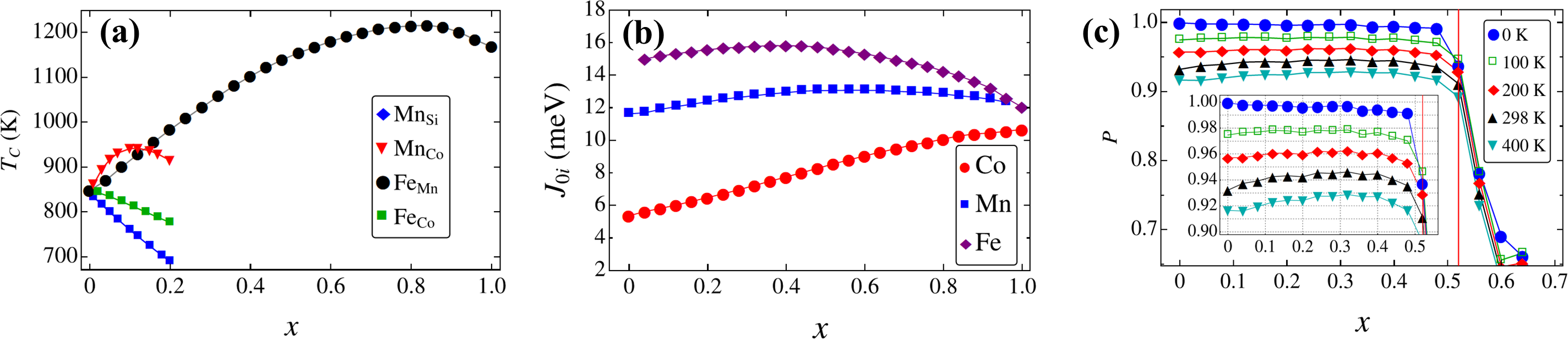}
	\caption{ (a) Mean-field Curie temperature $T_C$ for Co$_2$MnSi with Mn$_\mathrm{Si}$ or Mn$_\mathrm{Co}$ antisites and for the Co$_2$Mn$_{1-x}$Fe$_x$Si and (Co$_{1-x}$Fe$_x$)$_2$MnSi alloys. (b) Magnetic stability parameters $J_{0i}$ for Co, Mn, and Fe atoms in Co$_{2}$Mn$_{1-x}$Fe$_{x}$Si. (c) Spin polarization $P$ of the  DOS at the Fermi level in Co$_{2}$Mn$_{1-x}$Fe$_{x}$Si as a function of the concentration $x$ at several temperatures. The vertical red line at $x=0.52$ shows the point where the quasiparticle band with the Fe character crosses the Fermi level.}
		\label{fig.cmstcspn}
\end{figure*}

The effect of Mn$_{\textrm{Si}}$ and Mn$_{\textrm{Co}}$ antisites on the Curie temperature $T_C$ is shown in Fig. \ref{fig.cmstcspn}a. While $T_C$ decreases linearly with increasing concentration of Mn$_{\textrm{Si}}$, its dependence on the concentration of Mn$_{\textrm{Co}}$ is non-monotonic with a maximum at $x=0.12$. As for the vacancies, the Curie temperature is only slightly reduced by $\SI{53}{\kelvin}$ and $\SI{23}{\kelvin}$ in Co$_{2}$MnSi with a 5\% concentration of V$_{\mathrm{Co}}$ and V$_{\mathrm{Mn}}$, respectively.

We now examine the stability of the half-metallic gap in Co$_2$MnSi with Mn$_{\textrm{Si}}$ or Mn$_{\textrm{Co}}$ antisites at \emph{finite} temperatures. We note that Mn$_{\textrm{Sb}}$ antisites in a closely related half-metallic half-Heusler alloy NiMnSb are weakly exchange-coupled to their neighbors, which leads to strong spin disorder and loss of half-metallicity at relatively low temperatures \cite{NiMnSbpaper}. The magnetic stability of the given magnetic site $i$ is characterized by the parameter $J_{0i}=\sum_j J_{ij}$, where $J_{ij}$ is the pair exchange coupling.

The parameters $J_{0i}$ calculated using the linear response theory \cite{LIECHTENSTEIN} within CPA for Co$_2$MnSi with 15\% Mn$_{\textrm{Si}}$ or Mn$_{\textrm{Co}}$ antisites are listed in Table \ref{j0table}. We see that, in contrast with Mn$_\mathrm{Sb}$ antisites in NiMnSb, both types of Mn antisites in Co$_{2}$MnSi are strongly coupled to their neighbors. Therefore, thermal spin fluctuations are not expected to have a strong effect on the half-metallic gap at room temperature. Indeed, the spectral functions for Co$_{2}$MnSi with 15\% Mn$_{\textrm{Si}}$ or Mn$_{\textrm{Co}}$ antisites calculated at room temperature using the DLM method are similar to Figs.\ \ref{fig.defects}e and \ref{fig.defects}f and only show a small amount of additional broadening for some bands. Thus, we conclude that even large concentrations of Mn$_{\textrm{Si}}$ or Mn$_{\textrm{Co}}$ antisites do not strongly affect the half-metallic gap in Co$_2$MnSi either in the ground state or at room temperature.

\begin{table}[htb]
	\begin{tabular}{|l|l|c|}
		\hline
		\text{System} & \text{Atom} & \text{$J_0$ (\( \si{\milli\electronvolt} \))} \\
		\hline
		\multirow{2}{*}{Pure Co$_2$MnSi} & Co & 5.2\\ \cline{2-3}
		& Mn & 11.5\\  \cline{2-3}
		\hline
		\multirow{3}{*}{15\% Mn$_\mathrm{Si}$} & Co & 7.4\\	\cline{2-3}
		& Mn & 12.6 \\	\cline{2-3}
		& Mn$_{\textrm{Si}}$ & 31.9\\
		\hline
		\multirow{3}{*}{15\% Mn$_\mathrm{Co}$} & Co & 6.6\\ \cline{2-3}
		& Mn & 12.0 \\ \cline{2-3}
		& Mn$_{\textrm{Co}}$ & 9.7\\
		\hline
	\end{tabular}
	\caption{Magnetic stability parameters $J_{0i}$ for Co$_2$MnSi with Mn$_\mathrm{Si}$ and Mn$_\mathrm{Co}$ antisites.}
	\label{j0table}
\end{table}

\section{C\lowercase{o}$_2$M\lowercase{n}$_{1-x}$F\lowercase{e}$_x$S\lowercase{i} and (C\lowercase{o}$_{1-x}$F\lowercase{e}$_{x}$)$_2$M\lowercase{n}S\lowercase{i} alloys}

Alloying with Fe on the Mn sublattice shifts the Fermi energy in Co$_2$MnSi upward, which brings the Fermi level closer to the middle of the half-metallic gap, which is expected to be favorable for applications \cite{Fermitailoring_Fe,CFS_MTJ_TMR,CMS_E_struc,GMR_CMFS_MTJ}. In addition, alloying with Fe increases $T_C$ which is also favorable; in fact, Co$_{2}$FeSi has the highest $T_C$ of any Heusler alloy \cite{TCCFS}. In this section, we examine the evolution of the electronic structure of the CMFS alloy with increasing concentration $x$. We also consider the effects of the unavoidable Fe dopants on the Co sublattice.

\begin{figure*}[t]
	\centering
	\includegraphics[width=0.7\textwidth]{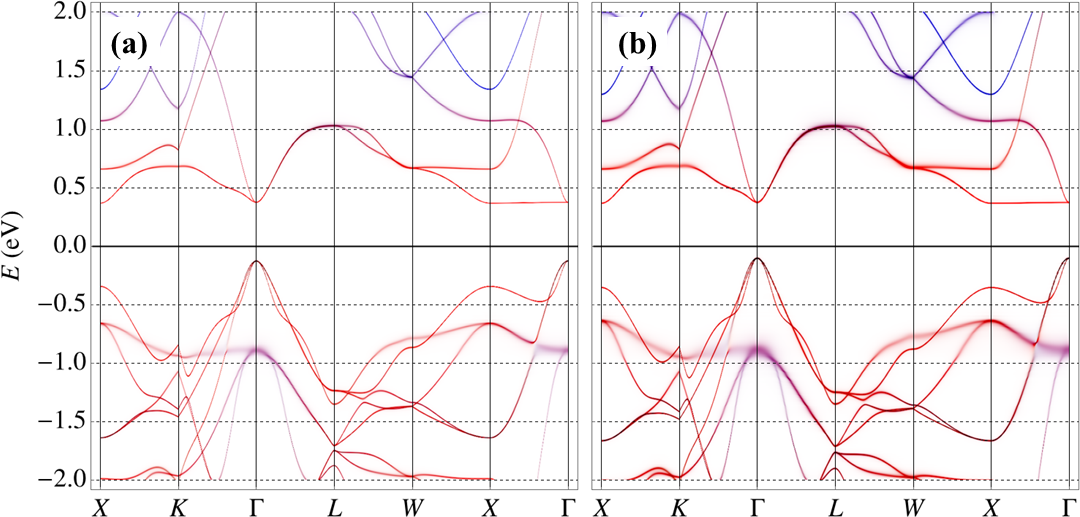}
	\caption{Minority-spin Bloch spectral functions in (Co$_{1-x}$Fe$_x$)$_2$MnSi at (a) $x=0.1$ and (b) $x=0.2$, both at $T=0$. The color scheme is the same as in Fig.\ \ref{fig.defects}.}
	\label{fig:COFE_alloy}
\end{figure*}

Figure \ref{fig.defects}(h-j) shows the minority-spin spectral functions of the CMFS alloy at $x=0.28$, 0.55, and 0.72, which can be compared with those for $x=0$ in Fig.\ \ref{fig.defects}a and $x=1$ in Fig.\ \ref{fig.defects}b. Because Fe has an extra valence electron compared to Mn, its addition increases the band filling, shifting the Fermi level upward. As seen in Fig.\ \ref{fig.defects}h, the Fermi level lies close to the middle of the gap at $x=0.28$. However, in addition to the band filling trend, the band structure also undergoes drastic changes with increasing $x$. The failure of the rigid-band model is already obvious from the fact that the band structure of the unoccupied states is very different in Co$_2$MnSi and Co$_2$FeSi; see Figs.\ \ref{fig.defects}a and \ref{fig.defects}b. The most important change in the band structure is the formation of a light band with the Fe character, which drops below the Mn-dominated unoccupied bands and crosses the Fermi level at the $X$ point at $x\approx0.52$, as seen in Fig.\ \ref{fig.defects}(h-j).

The magnetic stability parameters $J_{0i}$ for Co, Mn, and Fe atoms in CMFS are shown in Fig.\ \ref{fig.cmstcspn}b. At small concentrations of Fe, the exchange coupling of Mn and Fe atoms to their neighbors is similar and about twice stronger compared to Co. As $x$ increases from 0 to 1, the $J_{0,\mathrm{Co}}$ parameter increases by a factor of 2 and nearly reaches those of Mn and Fe, which pass through a broad maximum and then decline somewhat. As seen in Fig.\ \ref{fig.cmstcspn}a, the Curie temperature increases up to $x\approx0.8$ where it is about 40\% larger compared to Co$_2$MnSi. As seen in Fig.\ \ref{fig:MT}b, the enhancement of $J_{0,\mathrm{Co}}$ by alloying with Fe slightly increases the magnetic order parameter for Co while other sublattices are essentially unaffected.

Figure \ref{fig.cmstcspn}c shows the spin polarization $P=[N_\uparrow(E_F)-N_\downarrow(E_F)]/[N_\uparrow(E_F)+N_\downarrow(E_F)]$ of the density of states (DOS) $N_\sigma(E)$ at the Fermi level calculated using the DLM method in the range of temperatures from 0 to 400 K. The precipitous drop of the spin polarization at $x\approx0.52$ is caused by the light Fe band crossing the Fermi level. The deviation of $P$ from 1 at smaller concentrations of Fe reflects the effect of spin fluctuations without any quasiparticle bands appearing at the Fermi level. At room temperature $P$ starts at about 0.93 at $x=0$, increases almost to 0.95 at about $x\approx0.30$, and then declines back to 0.93 at $x=0.5$. This behavior is due to the competition of several factors: the upward shift of the Fermi level relative to the half-metallic gap, the increase in band broadening due to substitutional disorder, and the rearrangement of the band structure with the lowering of the light Fe band toward the Fermi level. Thus, there is an optimal concentration of about 30\% of Fe at which CMFS exhibits the maximum spin polarization at room temperature.

Fe atoms substituting on the Co sublattice in Co$_2$MnSi have a small magnetic moment of about 0.2 $\mu_B$. Such small magnetic moments should be regarded as being induced by the neighboring magnetic atoms, and the DLM spin fluctuation model is inapplicable to them. The minority-spin spectral functions of Co$_2$MnSi with 10 or 20\% of Co atoms substituted by Fe are shown in Fig.\ \ref{fig:COFE_alloy}. It is seen that Fe substitution for Co shifts the Fermi level downward toward the VBM but only slightly. This is because the reduction in band filling is accompanied by the reduction in the exchange splitting in the alloy where Fe is essentially non-magnetic. Aside from the small Fermi level shift, there are no substantial changes in the band structure. Electronic states near the $\Gamma$ point at about \SI{-1}{\eV} have a mixed Co-Fe and Mn character and are broadened by alloying with Fe, but they are far from the Fermi energy and do not affect the half-metallic gap, which is essentially unchanged even at $x=0.2$. Thus, partial substitution of Fe on the Co sublattice is not expected to have a significant effect on the spin polarization of the alloy, either at zero or at room temperature.

The Curie temperature is reduced by about \SI{70}{\kelvin} when 20\% Fe substitutes for Co. Thus, if a substantial fraction of Fe atoms substitutes on the Co sublattice, the increasing $T_C$ trend seen in Fig.\ \ref{fig.cmstcspn}a for CMFS may be weakened.

\section{Conclusions}

We have examined the electronic structure of Co$_2$MnSi in the presence of vacancies and Mn antisites, as well as those of the Co$_2$Mn$_{1-x}$Fe$_x$Si and (Co$_{1-x}$Fe$_x$)$_2$MnSi alloys, paying special attention to the effects of thermal spin fluctuations treated within the DLM approximation. Although small concentrations of Co or Mn vacancies do not destroy the half-metallic gap, they can lead to a considerable downward shift of the Fermi level, which crosses the VBM at about 5\% concentration. On the other hand, even large concentrations of Mn$_\mathrm{Si}$ and Mn$_\mathrm{Co}$ antisites do not strongly affect the half-metallic gap at room temperature. In the CMFS alloy, substitution of Mn by Fe shifts the Fermi level away from the VBM, but it also pulls down a light minority-spin band with the Fe character, which crosses the Fermi level and forms an electron pocket at about 50\% substitution. The spin polarization of DOS at the Fermi level has a broad maximum at about 30\% concentration at room temperature, increasing by about 0.02 from its value in pure Co$_2$MnSi. Fe atoms substituting on the Co sublattice are essentially non-magnetic; this substitution tends to shift the Fermi level slightly toward the VBM but otherwise does not affect the electronic states near the Fermi level.

\begin{acknowledgments}

We are grateful to Paul Crowell for useful discussions. This work was supported by the National Science Foundation (NSF) through Grant No.\ DMR-1609776 and the Nebraska Materials Research Science and Engineering Center (MRSEC, Grant No.\ DMR-1420645). Calculations were performed utilizing the Holland Computing Center of the University of Nebraska, which receives support from the Nebraska Research Initiative.

\end{acknowledgments}


\begin{thebibliography}{34}
\bibitem [{not()}]{note-radii}%
  \BibitemOpen
  \href@noop {} {}\bibinfo {note} {The sphere radii were set to 2.691 a.u. for
  Co, 2.6784 a.u. for Mn and Fe, and 2.4288 a.u. for Si; this results in the
  average overlap of 13.9\%.}\BibitemShut {Stop}%
\end{thebibliography}
\end{document}